\documentclass[aps,pra,reprint,nofootinbib,superscriptaddress,twocolumn,showpacs,showkeys,longbibliography,amsmath,amssymb]{revtex4-2}
\usepackage{graphicx}
\usepackage{dcolumn}
\usepackage{bm}
\usepackage{braket}
\usepackage{subfigure}
\usepackage[colorlinks,bookmarks=false,citecolor=blue,linkcolor=red,urlcolor=blue]{hyperref}
\usepackage[english]{babel}
\usepackage{changes}
\usepackage{multirow}
\begin{document}
\preprint{APS/123-QED}	
	
\title{Probing $p$-wave effects in spin-density separation of Bose mixtures with the dynamic structure factor}
\author{Xiaoran Ye}
\affiliation{Department of Physics, Zhejiang Normal University, Jinhua 321004, People's Republic of China}
\author{Zhaoxin Liang}\email[Corresponding author:~] {zhxliang@zjnu.edu.cn}
\affiliation{Department of Physics, Zhejiang Normal University, Jinhua 321004, People's Republic of China}
\date{\today}

\begin{abstract}
Quantum mixtures of Bose gases with tunable $s$- and $p$-wave interactions offer a versatile platform to explore strongly correlated phases and exotic phenomena. While repulsive interactions often drive phase separation, the interplay of 
$p$-wave interactions with spin-density decoupling remains underexplored. In this work, we employ the path integal field theory to investigate the role of  $p$-wave interactions in three-dimensional two-component Bose gas. We derive the Lee-Huang-Yang corrections to the ground-state energy and quantum depletion, revealing how 
$p$-wave interactions modify equations of state of the model system. Furthermore, we demonstrate that 
$p$-wave interactions predominantly renormalize the spin-sector effective mass in the language of decoupling the density and spin-density degrees of freedom. This effect manifests in the dynamic structure factor, computed via hydrodynamic theory, where Bragg spectroscopy can detect a tunable splitting between spin and density modes. Our results bridge theoretical predictions with experimental observables, offering insights into anisotropic interaction effects in quantum gases and their implications for probing emergent phases.
\end{abstract}
\maketitle

\section{Introduction}\label{INTRO}

At present, there is considerable and sustained interest in delving into the realm of $p$-wave pairing-dominated physics within the framework of quantum gas mixtures, distinguished by an orbital angular momentum of $l=1$~\cite{Gubbels2007}.
The motivation behind these interests is twofold. First, compared to $s$-wave pairing-dominated physics, pairing with finite angular momentum hosts richer structural complexity and plays a critical role in novel superconductors and superfluids~\cite{Luciuk2016}. This is exemplified by two landmark systems: (i) superfluid $^3$He where the $p$-wave pairing governs the symmetry-dependent emergence of gapless (A-phase) and gapped (B-phase) superfluidity~\cite{Vollhardt1990}, and (ii) Sr$_2$RuO$_4$, a proposed two-dimensional chiral $p_x+ip_y$ superconductor harboring Majorana edge modes with non-Abelian statistics---a cornerstone for topological quantum computing~\cite{Elliott2015, Read2000}.
 Second, advances in experimental techniques, notably Feshbach resonance~\cite{Regal2003,Viel2016,Zhang2004,Schunck2005,Gunter2005,Ospelkaus2006,Fuchs2008,Chevy2005,Liu2018,Venu2023}, enable precise tuning of 
$p$-wave interactions in ultracold gases. This unprecedented control over inter-intraspecies interactions revitalizes efforts to realize  $p$-wave superfluids while bridging insights to condensed-matter systems, such as unconventional superconductors~\cite{Read2000,Levinsen2007}.

Along this research line of pursuing novel quantum phases in quantum gas mixtures, a paramount prerequisite is the precise control of miscibility, governed by the interplay between atomic interactions and experimental geometry~\cite{Esry1997,Timmermans1998}. For $s$-wave systems, miscibility in binary Bose-Einstein condensates (BECs) is well-characterized across static~\cite{Zhou2008,Jiang2019,Zhan2014} and dynamical regimes~\cite{Wen2020,Gui2023}. In contrast, $p$-wave interactions introduce unique paradigms: mean-field studies predict their ability to regulate phase separation~\cite{Deng2024}, and notably, they can dramatically increase the effective mass of polaronic neutrons, causing it to diverge as the alpha condensate density approaches a critical threshold~\cite{Tajima2025}. Meanwhile, quantum corrections stabilize exotic states (e.g., polar droplets) and extend thermodynamics to higher partial waves~\cite{Gao2024,Li2019}. However, critical gaps persist in understanding the interplay between tunable $p$-wave and $s$-wave interactions---particularly their combined effects on non-equilibrium dynamics and experimental observables---posing both a theoretical and technical frontier.

Another impetus for this work originates from the measurement of spin-density (charge) separation in one-dimensional fermionic systems with tunable interactions, a phenomenon recently resolved via Bragg spectroscopy. In such systems, the dynamic structure factor (DSF) distinctly reveals two peaks corresponding to spin-density and density modes, as demonstrated in ultracold atomic experiments~\cite{Ruwan2022}. Beyond one-dimensional fermionic systems, Bragg spectroscopy has proven equally indispensable in higher-dimensional bosonic systems, where it not only quantifies interaction-driven effects but also provides a window into collective excitations and emergent many-body phenomena, as evidenced by studies spanning optical lattices to dipolar condensates~\cite{Enciso1995, Li2015, Hu2009}. Crucially, unlike one-dimensional Luttinger liquids requiring bosonization to decouple degrees of freedom~\cite{Chung2008}, spin-density separation in higher-dimensional bosonic systems arises intrinsically, thereby simplifying theoretical analyses by reducing interspecies correlations to intra-species correlations. This inherent decoupling enables both clearer experimental visualization and deeper insights into interaction mechanisms, as highlighted in quantum gas microscopy and momentum-resolved spectroscopy studies~\cite{Giamarchi2003, Recati2003, Kleine2008, Thomson2023}. At the Gaussian fluctuation level, spin-density separation further reveals that $p$-wave interaction effects in three-dimensional (3D) Bose mixtures predominantly manifest in the spin sector. These effects are experimentally accessible through Bragg spectroscopy, where the DSF exhibits a systematic shift between peaks under varying interaction strengths---a direct signature of spin-density interplay.

In this work, we are motivated to employ effective field theory within the one-loop approximation to derive analytical expressions for the ground-state energy and quantum depletion of a 3D Bose mixture with $p$-wave interactions at zero temperature.  By tuning the interaction strength, our results recover those for homogeneous Bose mixtures as presented in Refs.~\cite{Petrov2015, Chiquillo2018}. Beyond this, we formulate the effective action governing the decoupled density and spin-density degrees of freedom, and compute their corresponding ground-state energies. Furthermore, we present an expression for the DSF, elucidating how $p$-wave interactions induce spin-density separation---a phenomenon amenable to experimental verification. Remarkably, in the limit of vanishing $p$-wave interactions, our findings align with those of Ref.~\cite{Chung2008}. These results provide significant insights into the influence of anisotropic interactions on the properties of complex quantum many-body systems.

The paper is organized as follows. In Sec.~\ref{MODEL}, we introduce the action of the model under $p$-wave interactions within the framework of path integrals and derive the quasiparticle excitation spectrum using effective field theory under the one-loop approximation. In Sec.~\ref{EOS}, we calculate the analytical expressions for the ground-state energy and quantum depletion under $p$-wave interactions using the same effective field theory framework. In Sec.~\ref{DSF}, we explore the system's effective action under $p$-wave interactions and derive the ground-state energies of different degrees of freedom through spin-density separation, alongside the corresponding DSF. This section also discusses how spin-density separation under varying $p$-wave interactions can be experimentally observed. Finally, Sec.~\ref{CON} summarizes our findings and outlines the conditions necessary for the experimental realization of the proposed scenario.

\section{The model system of Bose mixtures with the tunable $s$- and $p$-wave interactions}\label{MODEL}

\subsection{Partition function of the system}
In this work, we investigate a 3D two-component Bose gas with particular emphasis on the interplay between conventional 
$s$-wave contact interactions and higher-order $p$-wave coupling---a regime recently made experimentally relevant through Feshbach resonance techniques~\cite{Venu2023,Deng2024}. 
To systematically explore this system, we employ the path-integral framework~\cite{Nagaosa2013,Ye2024,Yu2024,Zhang2024}, where the quantum statistical properties are encoded in the Euclidean partition function as follows
\begin{align}
	\mathcal{Z} = \int \mathcal{D}\left[\bar{\Psi},\Psi\right] \exp\left\{-\frac{S\left[\bar{\Psi},\Psi\right]}{\hbar}\right\}, \label{Partion}
\end{align}
with the action functional $S\left[\bar{\Psi},\Psi\right]= \int_{0}^{\beta\hbar}d\tau \int d^{3}{\bf r}\mathcal{L}$ in Eq. (\ref{Partion}). Here, the concrete form of the Lagrangian density $\mathcal{L}$ reads
\begin{widetext}
\begin{align}
	\mathcal{L}\left[\bar{\Psi},\Psi\right] &= \bar{\Psi}\left({\bf r},\tau\right)\left[\hbar\partial_{\tau}-\frac{\hbar^{2}}{2m}\nabla^{2}-\mu\right]\Psi\left({\bf r},\tau\right) - g_{p}\left| \Psi_1\left({\bf r},\tau\right)\nabla\Psi_2\left({\bf r},\tau\right) - \Psi_2\left({\bf r},\tau\right)\nabla\Psi_1\left({\bf r},\tau\right) \right|^2 \nonumber \\
	&\quad + \frac{g + g_{12}}{4}\left(\left|\Psi_1\left({\bf r},\tau\right)\right|^2 + \left|\Psi_2\left({\bf r},\tau\right)\right|^2\right)^2 + \frac{g - g_{12}}{4}\left(\left|\Psi_1\left({\bf r},\tau\right)\right|^2 - \left|\Psi_2\left({\bf r},\tau\right)\right|^2\right)^2. \label{action}
\end{align}
\end{widetext}
In Equation (\ref{action}), $\Psi\left({\bf r},\tau\right)=\left[\Psi_1,\Psi_2\right]^T$ (with $\bar{\Psi}\left({\bf r},\tau\right)=\left[\Psi^*_1,\Psi^*_2\right]$ denoting the conjugate fields) represents the two-component bosonic complex fields, which vary in both space ${\bf r}$ and imaginary time $\tau$. The first term represents the single-particle kinetics with atomic mass $m$ and chemical potential $\mu$, while $\beta=1/k_BT$ defines the inverse of the thermal energy scale with $k_B$ being the Boltzmann constant and $T$ denoting the temperature of the model system. 

The two terms in the second line of Eq.~(\ref{action}) embody the $s$-wave interaction energy of the model system. The $g=4\pi \hbar^2 a/m$ and $g_{12}=4\pi \hbar^2 a_{12}/m$ denote the intra- and interspecies coupling constant, respectively with $g\neq g_{12}$ in the view of the relevant experiments.  The $a$ and $a_{12}$ are the corresponding $s$-wave intra- and interspecies scattering lengths.  In the language of total density channel and spin-density channel labelled by $n_\rho=(|\Psi_1|^2+|\Psi_2|^2)/\sqrt{2}$ and $n_\sigma=(|\Psi_1|^2-|\Psi_2|^2)/\sqrt{2}$ respectively,  the first and second terms in the second line of Eq.~(\ref{action}) are referred as to the total density and spin-density energy induced by $s$-wave interatomic interaction. 

The emphasis and value of this work reside in the incorporation of the $p$-wave contribution into the Lagrangian density $\mathcal{L}$, as elucidated by the second term in the first line of Eq.~(\ref{action}). 
In more details, the $p$-wave interaction strength is introduced as $g_p = 2\pi \hbar^2 \nu_p / m$ with $\nu_p$ being the $p$-wave scattering length~\cite{Idziaszek2006,Idziaszek2009,Deng2024}. In Ref.~\cite{Venu2023}, it has been demonstrated that $\nu_p$ can be broadly tuned from the negative unitary limit to the positive unitary limit. When $\nu_p = 0$, the system's action simplifies to that analyzed in Refs.~\cite{Petrov2015,Deng2024}. 

\begin{table}[ht] 
	\renewcommand
	\arraystretch{1.0}  
	\caption{\label{table1} Ground state phase diagram of the two-component Bose mixture under variation of the coefficient $y$ and $z$,where negative parameters represent attractive interactions, while positive parameters indicate repulsive interactions.} 
	\begin{ruledtabular}  
		\begin{tabular}{ccc} 
			Interaction Range & Effects                                 & Reference             \\ \hline 
			$y < -1$, $z=0$          & Quantum droplet                         & Ref.~\cite{Petrov2015}  \\  \hline
			\multirow{2}{*}{y=0, $z<0$ and $z \ge 0$} & Quantum droplet under                  & \multirow{2}{*}{Ref.~\cite{Li2019}} \\ 
		& $p$-wave  interaction               &                       \\ \hline
			$0 < y < 1$, $z=0$      &    Spin-density separation              & Ref.~\cite{Chung2008}  \\  \hline
			\multirow{2}{*}{$0 < y < 1, z \geq 0$} &      Spin-density separation                 & \multirow{2}{*}{This work} \\ 
			& under $p$-wave interaction                &                       \\ \hline
	$y>0$, $z=0$    &   Mixed bubble phase & Ref.~\cite{Naidon2021}\\ \hline
		 $y>0$, $z<0$ &  Magnetoroton & Refs.~\cite{Andreev2020,Utesov2024}\\\hline
			$y > 1$ , $z=0$          & Phase separation                        & Ref.~\cite{Alex2002}   \\\hline
			\multirow{2}{*}{$y>1, z > 0$} &      Phase separation under                  & \multirow{2}{*}{Ref.~\cite{Deng2024}} \\ 
			&     $p$-wave interaction           &                      
			 
		\end{tabular} 
	\end{ruledtabular}  
\end{table} 

We restrict our analysis to the case where all intraspecies interactions are repulsive ($g>0$) and focus on the superfluid phase, characterized by the spontaneous breaking of the U(1) gauge symmetry for each component~\cite{Cappellaro2017}. Within the framework of effective field theory, the field is represented as $\psi_{i}\left({\bf r},\tau\right)=\sqrt{n_{0}}+\eta_{i}\left({\bf r},\tau\right)$, where $n_{0}=\left|\psi_{i}\right|^{2}$ denotes the 3D condensate density under the Bogoliubov approximation~\cite{Chiquillo2018,Pitaevskii2016}, and $\eta_{i}\left({\bf r},\tau\right)$ represents the fluctuation fields above the condensate. The mean-field plus Gaussian approximation is derived by expanding the action in Eq.~(\ref{action}) up to the second order in $\eta_{i}\left({\bf r},\tau\right)$ and $\eta_{i}^{*}\left({\bf r},\tau\right)$. The grand potential is then expressed as follows:
\begin{eqnarray}
	\Omega & = & \Omega_{0}+\Omega_{g}
\end{eqnarray}
with $\Omega_{0}\left(\mu,\sqrt{n_0}\right)/V =  -gn_{0}^{2}-g_{12}n_{0}^{2}$ being  the mean-field grand-potential and  $\Omega_{g}\left(\mu,\sqrt{n_{0}}\right)$ representing Gaussian fluctuations.
To ensure that  $\sqrt{n_0}$ genuinely minimizes the action and serves as a characteristic of the 3D BEC, the linear terms in the fluctuation fields must vanish within the action. The mean-field approximation is subsequently derived by minimizing $\Omega_0$, yielding the condition $\partial \Omega_0 / \partial \sqrt{n_0} = 0$. This, in turn, determines the chemical potential $\mu= gn_{0}+g_{12}n_{0}$.

Before proceeding, let us provide a brief overview of the ground state phase diagram of the Lagrangian density $\mathcal{L}$ in Eq.~(\ref{action}). In this end, we are motivated to introduce dimensionless parameters, reading $y = g_{12}/g$ and  $z=g_3n_0 = 2m g_{p} n_{0}/\hbar^2 $ respectively.  These parameters $y$ and $z$ capture the influences of $s$-wave and $p$-wave interactions, respectively, on the ground state phase diagram. Table~\ref{table1} summarizes the ground state phases for different values of $y$ and $z$.

In the case where $z=0$, corresponding to the absence of $p$-wave interaction, (i) a quantum droplet phase exists for $y<-1$, as first studied in Ref.~\cite{Petrov2015}; (ii) for $-1<y<1$, the system remains miscible, and within the range $0<y<1$, spin-density separation can occur, enabling further experimental exploration of interaction effects on the system~\cite{Deng2024, Chung2008}; (iii) when $y>0$, a mixed bubble phase emerges as discussed in Ref.~\cite{Naidon2021}; (iv) when $y>1$, the system exhibits phase separation~\cite{Alex2002}.

Upon introducing $p$-wave interactions, tuning $z$ gives rise to distinct physical phenomena in the two-component system, such as $p$-wave-affected quantum droplets~\cite{Li2019}, $p$-wave-affected phase separation~\cite{Deng2024}, $p$-wave-affected magnetoroton~\cite{Andreev2020,Utesov2024}, and $p$-wave-affected spin-density separation, which is the focus of this work. These effects further enrich the phase behavior of the system and open up new avenues for theoretical and experimental investigation. In Sec.~\ref{DSF}, we provide a comprehensive discussion of $p$-wave affected spin-density separation and present the corresponding theoretical framework for probing this phenomenon.

\subsection{Bogoliubov excitations}
In the preceding subsection, we outlined the system’s action and explored the effects within the two-component mixture arising from variations in the interaction coefficients $y$ and $z$. In this subsection, our objective is to derive the excitation spectrum from the Gaussian-order action. Focusing on the quadratic terms in $\eta_i$ and $\eta_i^*$ from Eq.~(\ref{action}), and to diagonalize them, we apply a Fourier transformation and subsequently introduce the Nambu space formalism~\cite{Armaitis2015}. This enables us to recast the quadratic part of the action into the desired form.
\begin{eqnarray}
S_{2}\left[\eta_{1}^{*},\eta_{1},\eta_{2}^{*},\eta_{2}\right] & = & -\frac{\hbar}{2}\sum_{\boldsymbol{k}\ne0,n}{\bf \Phi}_{\boldsymbol{k}n}\boldsymbol{G}_{\boldsymbol{k}n}^{-1}{\bf \Phi}_{\boldsymbol{k}n}^{\dagger},
\end{eqnarray}
with $\hbar\boldsymbol{k}$ representing the momentum and $n$ indexing the Matsubara frequencies $\omega_{n}=2\pi n/\hbar\beta$. The vector
\begin{eqnarray}
{\bf \Phi}_{\boldsymbol{k}n} & = & \left(\eta_{1\boldsymbol{k}n}^{*}, \eta_{1-\boldsymbol{k}n}, \eta_{2\boldsymbol{k}n}^{*}, \eta_{2-\boldsymbol{k}n}\right)
\end{eqnarray}
resides in the appropriate Nambu space, while the inverse Green’s function of the system, $\boldsymbol{G}_{\boldsymbol{k}n}^{-1}$, takes the form
\begin{eqnarray}
-\hbar\boldsymbol{G}_{\boldsymbol{k}n}^{-1} & = & \left(\begin{array}{cc}
-\hbar G_{B\boldsymbol{k}n}^{-1} & \hbar\Sigma_{12}\\
\hbar\Sigma_{12} & -\hbar G_{B\boldsymbol{k}n}^{-1}
\end{array}\right),\label{QGF}
\end{eqnarray}
in which $G_{B\boldsymbol{k}n}^{-1}$ and $\Sigma_{12}$ are $2\times2$ submatrices. (Hereafter, $2\times2$ matrices are denoted by uppercase letters, and $4\times4$ matrices by bold uppercase letters.) The diagonal submatrices correspond to the single-component inverse Green’s functions, given by
\begin{eqnarray}
-\hbar G_{B\boldsymbol{k}n}^{-1} & = & \left(\begin{array}{cc}
-i\hbar\omega_{n}+A^{\prime}+gn_{0} & gn_{0}\\
gn_{0} & i\hbar\omega_{n}+A^{\prime}+gn_{0}
\end{array}\right),\nonumber \\
\end{eqnarray}
with $A^{\prime}=\left(1+z\right)A=(1+g_{3}n_{0})\hbar^{2}k^{2}/2m$, $g_{3}=2mg_{p}/\hbar^{2}$ and $i\ne j$. The off-diagonal matrix in Eq.~(\ref{QGF}) is the self-energy due to the interspecies coupling, can be expressed as $\hbar\Sigma_{12}  =  \left(g_{12}-g_{3}An_{0}\right)I$.

The foundation of our path-integral formalism hinges on the meticulous construction of the matrix $\boldsymbol{M}$, a key theoretical construct that encapsulates the intricate interplay of $s$-wave and $p$-wave interactions within the system. This matrix, derived by setting $\omega_n = 0$ in the inverse Green’s function, acts as a crucial link between the microscopic dynamics of particle interactions and the emergent macroscopic properties of the system. By providing a structured representation of these interactions,  $\boldsymbol{M}$ serves as a pivotal tool for deriving the Bogoliubov modes $E_{\pm}$, which encapsulate the system’s excitation spectrum under varying interaction strengths:
\begin{widetext}
\begin{eqnarray}
\boldsymbol{M} & = & -\hbar\boldsymbol{G}_{\boldsymbol{k}0}^{-1}
  =  \left(
	\begin{array}{cccc}
		A+gn_{0}+g_{3}An_{0}, & gn_{0}, & n_{0}\left(g_{12}-g_{3}A\right), & n_{0}g_{12}\\
		gn_{0}, & A+gn_{0}+g_{3}An_{0}, & n_{0}g_{12} & n_{0}\left(g_{12}-g_{3}A\right)\\
		n_{0}\left(g_{12}-g_{3}A\right) & n_{0}g_{12} & A+gn_{0}+g_{3}An_{0}, & gn_{0}\\
		n_{0}g_{12} & n_{0}\left(g_{12}-g_{3}A\right) & gn_{0} & A+gn_{0}+g_{3}An_{0}
	\end{array}\right),\label{m}
\end{eqnarray}
\end{widetext}

\begin{figure}[hbtp] 
	\begin{centering} 
		\includegraphics[scale=0.7]{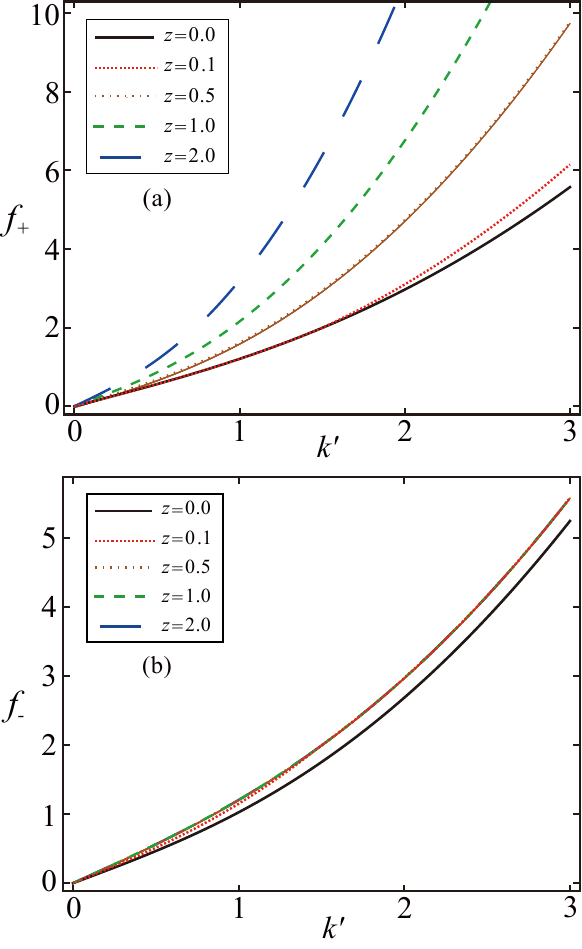} 
		\par\end{centering} 
	\caption{(a) Dimensionless excitation spectrum $f_{+}(k^{\prime})$ in Eq.~(\ref{ex1}) for different values of $z=g_{3}n_{0}$ from $0-2$. (b) Dimensionless excitation spectrum $f_{-}(k^{\prime})$ in Eq.~(\ref{ex1}) for different values of $z=g_{3}n_{0}$ from $0-2$. Here the dimensionless interaction parameter  $y=0.2$.\label{fig1}} 
\end{figure}
When the $p$-wave interaction strength $g_{p} = 0$, the contribution from $p$-wave scattering vanishes, leaving the interaction dynamics of the system governed solely by $s$-wave terms. In this limit, the matrix $\boldsymbol{M}$, which encapsulates the system’s interaction properties, reduces to a form consistent with the isotropic scattering configurations reported in Refs.~\cite{Armaitis2015,Chiquillo2018}, devoid of any anisotropic effects characteristic of $p$-wave interactions. Furthermore, when the interspecies interaction parameter $y = 0$, the coupling between different species disappears, effectively decoupling the system into two independent single-component Bose gases, and $\boldsymbol{M}$ aligns with the matrix derived in Ref.~\cite{Salasnich2016} for such a system. Central to our analysis, the Bogoliubov modes $E_{\pm}$, which characterize the system’s elementary excitations and provide essential insights into its stability and response to perturbations, are directly derived from $\boldsymbol{M}$ using the Cayley-Hamilton theorem, as detailed in Ref.~\cite{Wang2022}:
\begin{eqnarray}
E_{\pm} & = & \sqrt{\frac{1}{4}{\rm Tr}\left(\left(\boldsymbol{\kappa}\boldsymbol{M}\right)^{2}\right)\pm\sqrt{\frac{1}{16}{\rm Tr}\left(\left(\boldsymbol{\kappa}\boldsymbol{M}\right)^{2}\right)^{2}-\det\left(\boldsymbol{\kappa}\boldsymbol{M}\right)}},\nonumber \\
\label{ex}
\end{eqnarray}
these can be equivalently expressed as:
\begin{eqnarray}
E_{\pm} & = & gn_{0}f_{\pm}.\label{ex1}
\end{eqnarray}

Here, $f_{\pm}$ represent the two branches of the dimensionless excitation spectrum (see Appendix~\ref{AppendixA} for a detailed derivation). Fig.~\ref{fig1} provides a critical visualization of the Bogoliubov modes’ behavior, showcasing how $f_{\pm}$ evolve with the dimensionless wave vector $k^{\prime} = \hbar k / \sqrt{g n_0 m}$ as the $p$-wave interaction parameter $z$ varies from 0.1 to 2, with the interspecies interaction coefficient fixed at $y=0.2$. This figure underscores the profound influence of $p$-wave interactions, quantified by $z$, on the system’s excitation properties, revealing a transition from weakly perturbed $s$-wave-dominated modes to a regime where anisotropic $p$-wave scattering significantly reshapes the spectrum. Such insights are pivotal for understanding the interplay between interaction strengths and collective excitations in two-component Bose systems.

\section{Equation of state}\label{EOS}
\begin{figure}[t] 
	\begin{centering} 
		\includegraphics[scale=0.7]{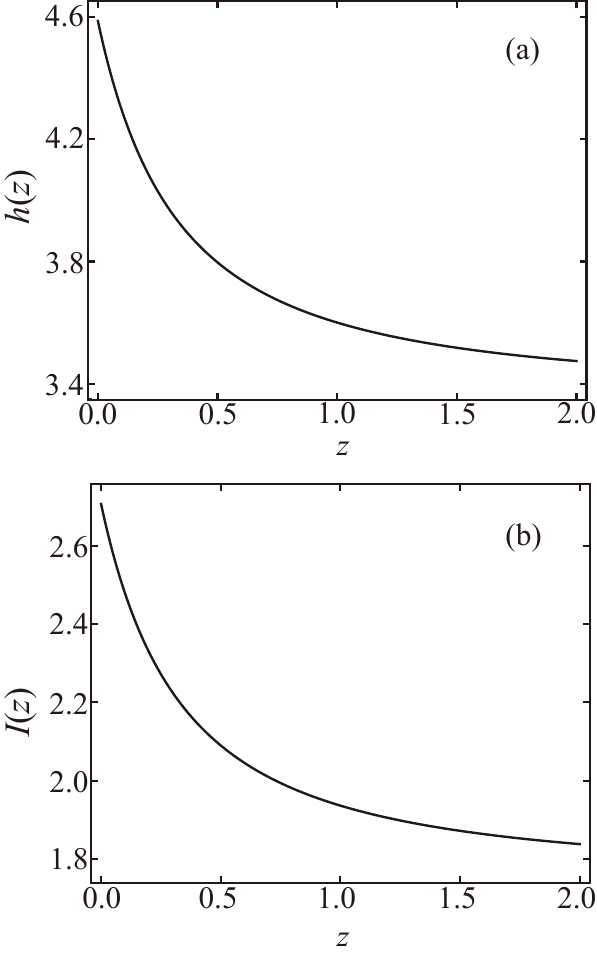} 
		\par\end{centering} 
	\caption{(a) Dimensionless scaling function $g(z)$ in Eq.~(\ref{GEF}) for different values of $z=g_{3}n_{0}$ from $0-2$. (b) Dimensionless scaling function $h(z)$ in Eq.~(\ref{QD}) for different values of $z=g_{3}n_{0}$ from $0-2$. Here the dimensionless interation parameters  $y=0.2$.\label{fig2}} 
\end{figure}

In Sec.~\ref{MODEL}, we introduced the Lagrangian density of the system and derived the excitation spectrum within the framework of effective field theory. In the subsequent Sec.~\ref{EOS}, our objective is to derive explicit analytical expressions for the Lee-Huang-Yang (LHY) corrections to the ground-state energy and quantum depletion of a 3D Bose mixture under the one-loop approximation. Note that the LHY correction to the ground-state energy of a Bose system with $s$-wave interactions has been previously derived using the one-loop approximation~\cite{Petrov2015, Chiquillo2018, Cappe12017,Tononi2018, Salasnich2017, Zhang2024,Ye2024,Yu2024}. In contrast, this work leverages the coherent-state path-integral formalism to systematically include $p$-wave interactions, offering a robust framework that overcomes the limitations of traditional perturbative methods. This approach facilitates a comprehensive analysis of the contributions from $p$-wave interactions to the ground-state properties, extending beyond the LHY correction for Bose mixtures with both $s$-wave and $p$-wave interactions. Our starting point is the grand potential of the Gaussian fluctuations, $\Omega_g\left(\mu, \sqrt{n_0}\right)$, expressed as:
\begin{eqnarray}
\Omega_{g}\left(\mu,\sqrt{n_{0}}\right) & = & \frac{1}{2}\sum_{\boldsymbol{k},\pm}\left[E_{\pm}+\frac{2}{\beta}\ln\left(1-e^{-\beta E_{\pm}}\right)\right],\nonumber \\
\end{eqnarray}
In this study, we primarily focus on the equation of state (EOS) of the model system at zero temperature. A key aspect of our analysis is to elucidate how $p$-wave interactions, parameterized by $g_{p}$, modify these LHY corrections, particularly through their influence on the excitation spectrum $E_{\pm}$. This $p$-wave contribution introduces momentum-dependent effects that distinguish the two-component mixture from purely $s$-wave-dominated systems. Notably, these effects also pave the way for exploring spin-density separation in the subsequent section, akin to observations in multi-component Bose systems. To achieve this, we calculate the ground-state energy of the system, given by $E_{\text{g}} = \Omega_{0} + \Omega_{g} + 2V \mu n_{0}$, as follows:
\begin{widetext}
\begin{eqnarray}
\frac{E_{\text{g}}}{V} & = & gn_{0}^{2}+g_{12}n_{0}^{2}+\frac{gn_{0}}{2V}\sum_{k\ne0}\left\{ f_{+}+f_{-}-\frac{\hbar^{2}k^{2}}{gn_{0}m}\left(1+z\right)-2+\frac{y^{2}+\left(y+1\right)^{2}z+1}{\left(2z+1\right)\frac{\hbar^{2}k^{2}}{2gn_{0}m}}\right\} ,\label{GSED}
\end{eqnarray}
\end{widetext}
In Eq.~(\ref{GSED}), the first and second terms on the right-hand side represent the mean-field contribution, whereas all subsequent terms account for the beyond-mean-field corrections stemming from quantum fluctuations. It is noteworthy that the last three terms in Eq. (\ref{GSED}) are introduced to remove power-law ultraviolet divergences in the momentum summation, a process that can be absorbed into an appropriate renormalization of the coupling constant $g_p$~\cite{braaten1997}. While this approach aligns with established techniques for handling divergences, as exemplified by $s$-wave renormalization in Ref.~\cite{Hu2020}, we refrain from an explicit derivation of the renormalized $g_p$ here. Instead, this subtraction ensures a finite LHY correction, consistent with the physical objectives of our study; the detailed derivation of these three terms is provided in Appendix \ref{AppendixB}. When we take the limit $z = 0$ or further set $y = 0$, Eq.~(\ref{GSED}) reduces to the form presented in Refs.~\cite{Chiquillo2018,Tononi2018}, double-checking the reasonableness and correctness of our approach. In the continuum limit, the summation in Eq.~(\ref{GSED}) can be systematically replaced by an integral, yielding the analytical expression for the ground-state energy of the model system as follows:
\begin{eqnarray}
\frac{E_{\text{g}}}{V} & = & gn_{0}^{2}+g_{12}n_{0}^{2}+\frac{1}{\left(2\pi\right)^{2}}\left(\frac{m}{\hbar^{2}}\right)^{3/2}\left(gn_{0}\right)^{5/2}h(z),\nonumber \\
\label{GSE}
\end{eqnarray}
where the scaling function $h\left(z\right)$ in terms of the variable $z$ can be solved analytically and the result reads
\begin{eqnarray}
h\left(z\right) & = & \frac{32}{30z+15}\left(\left(y+1\right)^{5/2}(1+2z)+\frac{(1-y)^{5/2}}{\sqrt{1+2z}}\right).\nonumber \\
\label{GEF}
\end{eqnarray}
When the parameter $y=0.2$, the corresponding result of Eq.~(\ref{GEF}) are presented in Fig.~\ref{fig2}(a). In the scenario of vanishing $p$-wave interaction, i.e., $z=0$, the ground-state energy of Eq.~(\ref{GSE}) reduces to
\begin{eqnarray}
\frac{E_{\text{g}}^{(0)}}{V} & = & gn_{0}^{2}+g_{12}n_{0}^{2}\nonumber \\
 &  & +\frac{1}{\left(2\pi\right)^{2}}\left(\frac{m}{\hbar^{2}}\right)^{3/2}\left(gn_{0}\right)^{5/2}h^{(0)}(y),\label{GS0E}
\end{eqnarray}
where the function of $h^{(0)}(y)$ reads
\begin{eqnarray}
h^{(0)}\left(y\right) & = & \frac{32}{15}\left(\left(y+1\right)^{5/2}+(1-y)^{5/2}\right),
\end{eqnarray}
and our result in Eq.~(\ref{GS0E}) can precisely recover the relevant ones in Ref.~\cite{Petrov2015}. Furthermore, when $y=0$, this expression reduces to the LHY correction for a single-component Bose system, aligning perfectly with the established single-species result in Ref.~\cite{Salasnich2016}.

Quantum depletion reflects the proportion of particles in nonzero momentum states, a characteristic we probe through the thermodynamic relation $N=-\partial \left(\Omega_{0}+\Omega_{g}\right)/\partial\mu$, with $\Omega_{0}+\Omega_{g}$ denoting the grand potential at zero temperature. The term $\Omega_{0}$ accounts for the mean-field contribution, whereas $\Omega_{g}$ incorporates the zero-temperature quantum fluctuations arising from Gaussian corrections. This framework yields the total particle density $n=n_{0}+\frac{1}{2V}\sum_{\boldsymbol{k},\pm}\partial_{\mu} E_{\pm}$, and by transforming summations into integrals within the path-integral approach, we arrive at the analytical form of the quantum depletion $\frac{N-N_{0}}{N}$ as follows:
\begin{eqnarray}
\frac{N-N_{0}}{N} & = & \frac{1}{(2\pi)^{2}}\left(\frac{m}{\hbar^{2}}\right)^{3/2}\left(g^{3}n\right)^{1/2}I\left(z\right),\label{QD}
\end{eqnarray}
where the scaling function $I\left(z\right)$ in terms of the variable $z$ is defined as
\begin{eqnarray}
I\left(z\right) & = & \frac{4}{3}\left(\frac{\left(1-y\right)^{3/2}}{\left(2z+1\right)^{3/2}}+\left(y+1\right)^{3/2}\right),\label{QDE}
\end{eqnarray}
when the parameter $y = 0.2$, the corresponding results of Eq.~(\ref{QDE}) are illustrated in Fig.~\ref{fig2}(b). To validate the accuracy of our findings, we consider the limit in which the $p$-wave interaction is absent. By setting $z = 0$, Eq.~(\ref{QD}) simplifies to the following form:
\begin{eqnarray}
\frac{N-N_{0}}{N} & = & \frac{1}{(2\pi)^{2}}\left(\frac{m}{\hbar^{2}}\right)^{3/2}\left(g^{3}n\right)^{1/2}I^{(0)}\left(y\right),\label{QD0}
\end{eqnarray}
where the scaling function $I^{(0)}\left(y\right)$ reads
\begin{eqnarray}
I^{(0)}\left(y\right) & = & \frac{4}{3}\left(\left(1-y\right)^{3/2}+\left(y+1\right)^{3/2}\right),\label{QDE0}
\end{eqnarray}
which is exactly consistent with the corresponding result in Ref.~\cite{Chiquillo2018}. Additionally, setting $y=0$ further simplifies the result, yielding the quantum depletion consistent with the result for a single-component Bose gas~\cite{Tononi2018}.

Before embarking on a detailed exploration of spin-density separation, it is essential to verify the physical plausibility of the dimensionless parameters $y = g_{12}/g$ and $z = g_3 n_0 = \frac{2m}{\hbar^2} g_p n_0$ used in our computations of ground-state energy and quantum depletion.  This validation is critical, as it establishes the experimental feasibility of our theoretical framework for a two-component Bose system. The value of $y$ can be determined by the Feshbach resonance in a dual-species system, taking Ref.~\cite{Wang2016} as an example, we set $a=54.5a_{\text{B}}$, and $a_{12}$ can be tuned from $-28a_\text{B}$ to $502a_\text{B}$ via an interspecies Feshbach resonance, enabling $y$ to range from $-0.51$ to $9.21$. Meanwhile, the range of $z$ can be informed by Ref.~\cite{Venu2023}, where the $p$-wave scattering volume $\nu_{p}$ is tuned via a Feshbach resonance from approximately 0 to unitary limit, leading to the actual range of $z$ far exceeding the interval of 0 to 2. These ranges for $y$ and $z$ are consistent with the parameter space analyzed in our ground-state energy and quantum depletion studies, as depicted in Figs.~\ref{fig1},~\ref{fig2}, and following Fig.~\ref{fig3}.

Eqs.~(\ref{GSE}) and~(\ref{QD}) constitute the central results of this study, offering explicit analytical expressions for the LHY order corrections to the ground-state energy and quantum depletion under the influence of $p$-wave interactions. Both equations exhibit a high-order decay behavior with respect to the coefficients of the $p$-wave interactions. In the following analysis, we will explore the underlying mechanisms responsible for this high-order decay from the perspective of spin-density separation. Inspired by Ref.~\cite{Chung2008}, we provide theoretical predictions for the experimental observation of $p$-wave-induced spin-density separation, anticipating a tunable splitting in the structure factor due to the variable particle-number fluctuations in the spin degrees of freedom induced by $p$-wave interactions.

\section{PROBING SPIN-DENSITY SEPARATION USING BRAGG SPECTROSCOPY}\label{DSF}

In the preceding Sec.~\ref{EOS}, we derived explicit analytical expressions for the LHY-order corrections to the ground-state energy and quantum depletion of a 3D Bose gas within the one-loop approximation. The objective of Sec.~\ref{DSF} is to analyze the effects of $p$-wave interactions on the system from the perspective of spin-density separation and to propose an experimental scheme to observe this phenomenon through calculations of the system's DSF.

To gain deeper insight into the low-energy excitations of Bose mixtures featuring $p$-wave interactions, we formulate an effective hydrodynamic Lagrangian at low energies, capturing solely the modes tied to these excitations. To this end, the boson field $\psi_{i}$ is decomposed into the number density $n_{i}$ and the phase $\theta_{i}$, expressed as $\psi_{i}=\sqrt{n_{i}}e^{i\theta_{i}}=\sqrt{n_{0}+\delta n_{i}}e^{i\theta_{i}}$, with $\delta n_i$ represents the density fluctuation. By adopting a transformed basis defined by $\delta n_{\rho\left(\sigma\right)}=\left(\delta n_{1}\pm\delta n_{2}\right)/\sqrt{2}$ and $\theta_{\rho\left(\sigma\right)}=\left(\theta_{1}\pm\theta_{2}\right)/\sqrt{2}$, the Gaussian order of the action in Eq.~(\ref{action}) in real time can be recast as follows:
\begin{eqnarray}
S_{g}\left(\boldsymbol{r},t\right) & = & \int d^{3}\boldsymbol{r}dt\sum_{\lambda=\rho,\sigma}\frac{n_{0}\left(\nabla_{r}\theta_{\lambda}\right)^{2}}{2m_{\lambda}}+\delta n_{\lambda}\partial_{t}\theta_{\lambda}\nonumber \\
 &  & +\delta n_{\lambda}\frac{1}{2}\left(\frac{\nabla_{r}^{2}}{4m_{\lambda}n_{0}}+g_{\lambda}\right)\delta n_{\lambda},\label{GA}
\end{eqnarray}
where the coupling strength $g_{\rho(\sigma)}=g(1\pm y)$, the density mass $m_{\rho}=m$, and the spin mass $m_{\sigma}=m/\left(1+z\right)$ account for effect of $p$-wave interaction. When $z = 0$, the effective action reduces to that presented in Ref.~\cite{Chung2008}. At the Gaussian level, it is apparent that the $p$-wave interaction is fully decoupled into the spin-density degree of freedom, effectively resulting in a renormalization of the effective mass in the kinetic energy term. Upon performing a Fourier transformation, the action $S_g\left(Q\right)$ assumes the following form
\begin{eqnarray}
S_{g}\left(Q\right) & = & \boldsymbol{\Psi}\left(Q\right)\boldsymbol{M}^{\prime}\boldsymbol{\Psi}^{\dagger}\left(-Q\right)
\end{eqnarray}
where $Q=\left(\boldsymbol{k},\omega\right)$ is $3+1$ vector denoting the momenta $\boldsymbol{k}$ and the frequency $\omega$, and
\begin{eqnarray}
\boldsymbol{\Psi}\left(Q\right) & = & \left(\begin{array}{cccc}
\delta n_{\rho}\left(Q\right), & \theta_{\rho}\left(Q\right), & \delta n_{\sigma}\left(Q\right), & \theta_{\sigma}\left(Q\right)\end{array}\right),\nonumber \\
\end{eqnarray}
is the vector that contains all the basis $\delta n_{\lambda}$, $\theta_{\lambda}$. What's more, the matrix $\boldsymbol{M}^{\prime}$ reads
\begin{eqnarray*}
\boldsymbol{M}^{\prime} & = & \left(\begin{array}{cc}
M_{\rho} & 0\\
0 & M_{\sigma}
\end{array}\right)
\end{eqnarray*}
with the $2\times2$ submatrices being
\begin{eqnarray}
M_{\rho,\sigma} & = & \left(\begin{array}{cc}
\frac{\hbar^{2}k^{2}}{8m_{\rho,\sigma}n_{0}}+\frac{g_{\rho,\sigma}}{2} & \frac{1}{2}i\omega_{\rho,\sigma}\\
-\frac{1}{2}i\omega_{\rho,\sigma} & \frac{\hbar^{2}n_{0}k^{2}}{2m_{\rho,\sigma}}
\end{array}\right),
\end{eqnarray}
the form of two submatrices is identical to that of the one-component inverse Green function. Consequently, we note that, in contrast to one-dimensional Fermi systems, which necessitate bosonization to decouple the density and spin-density degrees of freedom, the analysis here is streamlined by focusing solely on fluctuations around the condensate density~\cite{Chung2008}. This inherent decoupling substantially simplifies the theoretical framework while preserving a clear distinction between the density and spin-density modes.
\begin{figure*}[t]
	\begin{centering} 
		\includegraphics[scale=0.7]{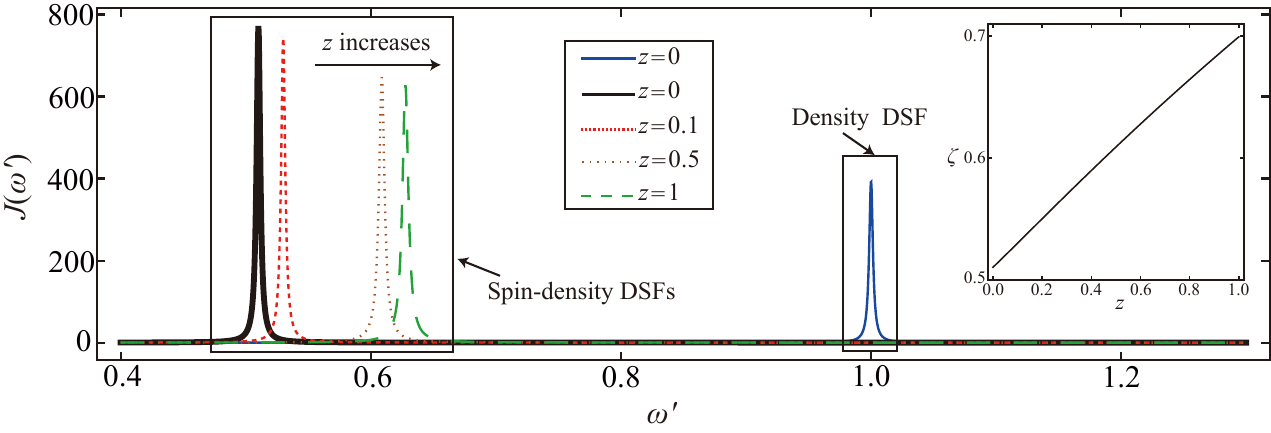} 
		\par\end{centering} 
	\caption{Dimensionless dynamic structure factor function $J(\omega^{\prime})$ in Eqs.~(\ref{J1}) and (\ref{J2}) in the hydrodynamic regime and at zero temperature for $y=0.5$ and $k^{\prime}=1.25$ shows different distinct peaks from $z=0-1$ corresponding to the density and the spin waves, centered at $1$ and $\zeta$. Inset: Dependence of $\zeta$ on $z$, illustrating the $p$-wave-induced spin-density separation.\label{fig3}} 
\end{figure*}

Then we can obtain two branches of excitations
\begin{eqnarray}
\varepsilon_{\rho\left(\sigma\right)}\left(k\right) & = & \sqrt{\frac{\hbar^{2}k^{2}}{2m_{\rho\left(\sigma\right)}}+\frac{\hbar^{2}g\left(1\pm y\right)n_{0}}{m_{\rho\left(\sigma\right)}}k^{2}},\label{ES}
\end{eqnarray}
where $\varepsilon_{\rho}\left(k\right)$ corresponds to the density mode and $\varepsilon_{\sigma}\left(k\right)$ corresponds to the spin-density mode. Using these excitation spectra, we can compute the ground-state energy of the density and spin-density degrees of freedom
\begin{eqnarray}
E_{g\rho\left(\sigma\right)} & = & \frac{g_{\rho\left(\sigma\right)}N^{2}}{2V}+\frac{8g_{\rho\left(\sigma\right)}^{5/2}N^{5/2}m_{\rho\left(\sigma\right)}^{3/2}}{15\pi^{2}\hbar^{3}V^{3/2}}.
\end{eqnarray}

We now explore how spin-density separation in a two-component Bose gas can be experimentally observed using Bragg spectroscopy. This technique entails inducing a density perturbation in the system while simultaneously probing both spin-density and density excitations. By employing two Bragg laser beams with momenta $\boldsymbol{k}_1$ and $\boldsymbol{k}_2$, and precisely adjusting their frequency difference $\omega$ (maintained significantly smaller than their detuning from the atomic resonance), it is feasible to selectively excite both the density and spin-density waves within the Bose gas~\cite{Li2015, Ernst2010, Du2010}.To quantitatively analyze these excitations, we turn to the DSF of these waves for $\omega > 0$, which can be approximated as follows~\cite{Chung2008}:
\begin{eqnarray}
S_{\rho\left(\sigma\right)}\left(k,\omega\right) & \approx & \frac{\chi_{\rho\left(\sigma\right)}v_{\rho\left(\sigma\right)}k\Gamma_{\rho\left(\sigma\right)}}{2\left[\left(\omega-v_{\rho\left(\sigma\right)}k\right)^{2}+\Gamma_{\rho\left(\sigma\right)}^{2}\right]},
\end{eqnarray}

Here, the ground-state compressibility $\chi_{\rho(\sigma)}$ is related to the ground-state energy $E$ through the expression $\chi^{-1}=\frac{1}{V}\frac{\partial^{2}E}{\partial n^{2}}$, where $V$ denotes the constant system volume and $n = N/V$ represents the density. The compressibility for 
\begin{align}
	\chi_{\rho}^{-1}=g_{\rho}+\frac{2g_{\rho}^{5/2}n_{0}^{1/2}m_{\rho}^{3/2}}{(\pi^{2}\hbar^{3})},
\end{align}
and the compressibility for the spin-density degree of freedom is given by
\begin{align}
	\chi^{-1}_{\sigma} &	= g_{\sigma} + \frac{2 g_{\sigma}^{5/2} m^{3/2} n_{0}^{1/2} z^{2}}{\pi^{2} \hbar^{3} (z+1)^{7/2}} \notag \\	
	&\quad - \frac{4 g_{\sigma}^{5/2} m^{3/2} n_{0}^{1/2} z}{\pi^{2} \hbar^{3} (z+1)^{5/2}} + \frac{2 g_{\sigma}^{5/2} m^{3/2} n_{0}^{1/2}}{\pi^{2} \hbar^{3} (z+1)^{3/2}}.
\end{align}
The sound velocity is similarly determined as $v_{\rho\left(\sigma\right)}=\left(\frac{V}{mn}\frac{\partial^{2}E}{\partial V^{2}}\right)^{1/2}$ with a constant particle number $N$, expressed for the density degree of freedom as
\begin{align} 
 v_{\rho}=\left(\frac{g_{\rho}n_{0}}{m_{\rho}}+\frac{2g_{\rho}^{5/2}m_{\rho}^{1/2}n_{0}^{3/2}}{\pi^{2}\hbar^{3}}
\right)^{1/2},
\end{align}
 and for the spin-density degree of freedom as
\begin{align}
	v_{\sigma} &= \left( \frac{2 g_{\sigma}^{5/2} m^{1/2} n_{0}^{3/2} z^{2}}{\pi^{2} \hbar^{3} (z+1)^{5/2}} + \frac{2 g_{\sigma}^{5/2} m^{1/2} n_{0}^{3/2}}{\pi^{2} \hbar^{3} (z+1)^{1/2}} \right. \notag \\
	& \quad \left. - \frac{4 g_{\sigma}^{5/2} m^{1/2} n_{0}^{3/2} z}{\pi^{2} \hbar^{3} (z+1)^{3/2}} + \frac{g_{\sigma} n_{0} (1+z)}{m} \right)^{1/2}.
\end{align}
Additionally, in calculating the damping rates $\Gamma_{\rho,\sigma}$, we follow Ref.~\cite{Chung2008} by introducing the Beliaev damping formalism developed for a single-component $s$-wave Bose gas, which determines the width of the DSF peaks. We assume that the $p$-wave interaction effects are primarily captured by modifying the spin-density mode’s effective mass $\left(m_\sigma=m/\left(1+z\right)\right)$, while the density mode retains the bare mass $m$, as derived in Eq.~(\ref{GA}). Based on this assumption, we approximate the damping rates as $\Gamma_{\rho\left(\sigma\right)}=3\hbar k^{5}/\left(640\pi m_{\rho\left(\sigma\right)}n_{0}\right)$, incorporating the respective effective masses for the density and spin-density modes. When the $p$-wave interaction vanishes $\left(z=0\right)$, the damping rate reduces to the $s$-wave form, consistent with Refs.~\cite{Chung2008,Chung2009}. This $k^5$ dependent form arises from low-energy quasiparticle scattering, with $p$-wave influences indirectly reflected through the mass terms. Importantly, our study focuses on the spin-density separation, driven by the energy splitting between the spin and density modes as given by the excitation energies $\varepsilon_{\rho}$ and $\varepsilon_{\sigma}$ (Eq.~(\ref{ES})), while  $\Gamma_{\rho,\sigma}$ affects only the broadening of the DSF peaks without significantly altering this separation. Using these damping rates, we reexpress the DSFs as follows:
\begin{eqnarray}
S_{\rho\left(\sigma\right)}\left(\omega\right) & = & \chi_{\rho}J_{\rho\left(\sigma\right)}\left(\omega^{\prime}\right),
\end{eqnarray}
where the dimensionless DSF function in density degree of freedom $J_{\rho}\left(\omega^{\prime}\right)$ reads
\begin{eqnarray}
J_{\rho}\left(\omega^{\prime}\right) & = & \frac{\Gamma_{\rho}/v_{\rho}k^{\prime}}{\left(\omega^{\prime}-1\right)^{2}+\left(\Gamma_{\rho}/v_{\rho}k^{\prime}\right)^{2}},\label{J1}
\end{eqnarray}
and in spin-density degrees of freedom
\begin{eqnarray}
J_{\sigma}\left(\omega^{\prime}\right) & = & \frac{\alpha\Gamma_{\sigma}/v_{\rho}k^{\prime}}{\left(\omega^{\prime}-\zeta\right)^{2}+\left(\Gamma_{\sigma}/v_{\rho}k\right)^{2}}.\label{J2}
\end{eqnarray}

Here, firstly, the momentum$k$ is rendered dimensionless as discussed in Sec.~\ref{MODEL}, where $k^\prime=\hbar k/\sqrt{g n_0 m}$, and here we set $\hbar/\sqrt{gn_0m}=1$. $\omega' = \omega / v_{\rho} k^{\prime}$ represents the dimensionless frequency, with additional dimensionless coefficients defined as $\alpha = \chi_{\sigma} v_{\sigma} / \chi_{\rho} v_{\rho}$ and $\zeta = v_{\sigma} / v_{\rho}$, where the dependence of $\zeta$ on $z$ is illustrated in the inset of Fig.~\ref{fig3}. Setting $k^{\prime} = 1.25$ and $y = 0.5$, the density compressibility $\chi_{\rho}$ can be approximately regarded as constant. The dimensionless DSF functions $J_{\rho\left(\sigma\right)}$ reflect the variation with frequency $\omega^{\prime}$ under different values of $z$, as depicted in Fig.~\ref{fig3}. From Fig.~\ref{fig3}, we observe that the density wave typically exhibits a higher sound velocity and a broader peak due to Beliaev damping, whereas the spin wave tends to display a narrower and more distinct response. Notably, the peak of the spin-density wave progressively shifts toward higher frequencies as the strength of the $p$-wave interaction increases. This offers a clear visualization of the influence of $p$-wave interactions on the system, providing a direct experimental signature detectable through Bragg scattering experiments.
\section{CONCLUSION}\label{CON}
The primary contribution of this article lies in deriving the EOS under $p$-wave interactions and elucidating their influence on a two-component Bose-Bose mixture through spin-density separation. Beyond providing explicit expressions for the LHY-order corrections to the ground-state energy and quantum depletion, as detailed in Sec.~\ref{EOS}, this work offers a visualization scheme via the DSFs and a comprehensive analysis of the theoretical results, as presented in Sec.~\ref{DSF}. A key finding is that $p$-wave interactions, parameterized by $z$, induce a tunable splitting in the DSFs, reflecting affected particle-number fluctuations in the spin degrees of freedom, which distinguishes this system from purely $s$-wave-dominated mixtures. Furthermore, the validity of the Bogoliubov approximation employed in our calculations is retrospectively confirmed by estimating the quantum depletion $\left(N-N_{0}\right)/N$, utilizing Fig.~\ref{fig2} (b). In typical Bose-Bose mixture experiments, with parameters $n \approx 3\times10^{14}~\text{cm}^{-3}$, $a\approx2.7517~\text{nm}$~\cite{Knoop2011}, the quantum depletion in Eq.~(\ref{QD}) is evaluated as $\left(N-N_{0}\right)/N\approx0.0028\times I \left(z\right) $ shown in Fig.~\ref{fig2} (b), affirming the applicability of the Bogoliubov approximation~\cite{Xu2006}.In addition, we briefly address the applicability of our single-channel model in the context of $p$-wave interactions, particularly near Feshbach resonances where a two-channel model might be more appropriate due to the coupling between open and closed channels. Our analysis employs a single-channel framework with the $p$-wave coupling $g_{p}=2\pi\hbar^{2}\nu_{p}/m$, tuned via the scattering volume $\nu_{p}$, which is sufficient to capture the weak-interaction regime ($0<y<1$ , $z$ from $0$ to $2$) explored in this work. This regime, far from the strong-coupling limit of $p$-wave Feshbach resonances, aligns with experimental conditions where $\nu_{p}$ remains moderate and single-channel descriptions hold, as valid ated by the tunability demonstrated in Ref.~\cite{Venu2023}. While a two-channel model could provide a more comprehensive description near resonance, enhancing predictions of resonance-enhanced effects, our focus on the Gaussian-level LHY corrections and spin-density separation justifies the simplicity and efficacy of the single-channel approach within the specified parameter space. 

The experimental realization of our scenario relies on precise control of three key parameters: the $s$-wave inter- and intra-species scattering lengths ($a_{12}$ and $a$) and the $p$-wave scattering volume ($v_{p}$), all of which are highly tunable via Feshbach resonance~\cite{Wang2016, Venu2023}.  We anticipate that these predicted features, particularly the $p$-wave-induced spin-density separation observable through Bragg spectroscopy, will be validated in future experiments. However, the treatment of Beliaev damping under $p$-wave interactions remains an open question deserving further study. While we have approximated the damping rates using an $s$-wave formalism with effective mass adjustments, a detailed analysis incorporating $p$-wave-specific scattering processes could refine our understanding of the DSF peak broadening and its experimental implications.

In summary, this paper examines the effects of $p$-wave interactions on the EOS and spin-density separation within a Bose-Bose mixture. At the Gaussian order, we demonstrate that $p$-wave interactions fully decouple the spin-density degree of freedom, resulting in a compression effect confined to its effective mass, as evidenced by the renormalization captured in Sec.~\ref{DSF}. This decoupling, driven by the momentum-dependent nature of $p$-wave interactions, not only manifests as distinct DSF peaks but also enriches the system’s phase behavior beyond traditional $s$-wave frameworks. However, the treatment of Beliaev damping under $p$-wave interactions remains an open question deserving further study. While we have approximated the damping rates using an $s$-wave formalism with effective mass adjustments, a detailed analysis incorporating $p$-wave-specific scattering processes could refine our understanding of the DSF peak broadening and its experimental implications. By bridging theoretical predictions with experimental feasibility, this study enhances our understanding of interaction-driven phenomena in multi-component Bose systems and provides robust theoretical guidance for future explorations in higher-dimensional quantum gases, potentially advancing the study of exotic quantum states influenced by $p$-wave interactions.

\section{ACKNOWLEDGMENTS}

 We thank Yi Zhang, Tao Yu, Ying Hu, and Biao Wu for stimulating discussions and useful help. This research was supported by  the Zhejiang Provincial Natural Science Foundation of China under Grant No. LZ25A040004 and the National Natural Science Foundation of China under Grant No. 12074344.
 
\onecolumngrid
\section*{Appendixes}
\appendix
\section{CAYLEY-HAMILTON THEOREM}\label{AppendixA}
The purpose of Appendix \ref{AppendixA} is to give the detailed derivation of Eq.~(\ref{ex1}) in the main text. 
From the Cayley-Hamilton theroem, we can get the excitation Eq.~(\ref{ex}) with Eq.~(\ref{m}), the matrix $\boldsymbol{M}$, which reads:
\begin{eqnarray}
\boldsymbol{M} & = & \left(\begin{array}{cccc}
A+gn_{0}+g_{3}An_{0}, & gn_{0}, & n_{0}\left(g_{12}-g_{3}A\right), & n_{0}g_{12}\\
gn_{0}, & A+gn_{0}+g_{3}An_{0}, & n_{0}g_{12} & n_{0}\left(g_{12}-g_{3}A\right)\\
n_{0}\left(g_{12}-g_{3}A\right) & n_{0}g_{12} & A+gn_{0}+g_{3}An_{0}, & gn_{0}\\
n_{0}g_{12} & n_{0}\left(g_{12}-g_{3}A\right) & gn_{0} & A+gn_{0}+g_{3}An_{0}
\end{array}\right),
\end{eqnarray}
where $A=\frac{\hbar^{2}k^{2}}{2m}$, $g_{3}=\frac{2m}{\hbar^{2}}g_{p}$, and 
$\boldsymbol{\kappa}=\left(
\begin{array}{cccc}
1 & 0 & 0 & 0\\
0 & -1 & 0 & 0\\
0 & 0 & 1 & 0\\
0 & 0 & 0 & -1
\end{array}\right)$.
We will now present the detailed derivation process of Eq.~(\ref{ex}) and Eq.~(\ref{ex1}), including a demonstration of how the excitation spectrum is nondimensionalized. Our starting point begins with the trace term in Eq.~(\ref{ex}), which takes form as follows:
\begin{eqnarray}
\frac{1}{4}{\rm Tr}\left(\left(\boldsymbol{\kappa}\boldsymbol{M}\right)^{2}\right) & = & (A\left(g_{3}n_{0}+1\right)+gn_{0})^{2}+n_{0}^{2}(g_{12}-Ag_{3})^{2}-g^{2}n_{0}^{2}-g_{12}^{2}n_{0}^{2}\nonumber \\
 & = & A\left(2g_{3}n_{0}^{2}(Ag_{3}+g-g_{12})+2n_{0}(Ag_{3}+g)+A\right),
\end{eqnarray}
By nondimensionalizing it, we obtain:
\begin{eqnarray}
\frac{{\rm Tr}\left(\left(\boldsymbol{\kappa}\boldsymbol{M}\right)^{2}\right)}{4\left(gn_{0}\right)^{2}} & = & B\left(2z(Bz+1-y)+2(Bz+1)+B\right)\\
 & = & B^{2}\left(2z^{2}+2z+1\right)+B\left(2z-2yz+2\right),
\end{eqnarray}
with $B=\frac{\hbar^{2}k^{2}}{2mgn_{0}}$, $z=g_{3}n_{0}$ and $y=g_{12}/g$. Then we can calculate the determinant term, which reads
\begin{eqnarray}
\det\left(\boldsymbol{\kappa}\boldsymbol{M}\right) & = & A^{2}\left(A+2n_{0}\left(g+g_{12}\right)\right)\left(A\left(2g_{3}n_{0}+1\right)+2gn_{0}-2g_{12}n_{0}\right)\left(2g_{3}n_{0}+1\right),\nonumber \\
\end{eqnarray}
and with its dimensionless form
\begin{eqnarray}
\frac{\det\left(\boldsymbol{\kappa}\boldsymbol{M}\right)}{\left(gn_{0}\right)^{4}} & = & B^{2}\left(B+2\left(1+y\right)\right)\left(B\left(2z+1\right)+2-2y\right)\left(2z+1\right),
\end{eqnarray}
So Eq.~(\ref{ex}) can be rewritten as
\begin{eqnarray}
E_{\pm} & = & gn_{0}\sqrt{\frac{{\rm Tr}\left(\left(\boldsymbol{\kappa}\boldsymbol{M}\right)^{2}\right)}{4\left(gn_{0}\right)^{2}}\pm\sqrt{\frac{{\rm Tr}\left(\left(\boldsymbol{\kappa}\boldsymbol{M}\right)^{2}\right)^{2}}{16\left(gn_{0}\right)^{4}}-\frac{\det\left(\boldsymbol{\kappa}\boldsymbol{M}\right)}{\left(gn_{0}\right)^{4}}}}\nonumber \\
 & = & gn_{0}f_{\pm},
\end{eqnarray}
which are exactly Eq.~(\ref{ex1}).
\section{REMOVING POWER ULTRAVIOLET DIVERGENCES OF EQ.~(\ref{GSED})}\label{AppendixB}
In Appendix \ref{AppendixB}, we give the detailed derivations of the crucial regularizing terms on the second line of Eq.~(\ref{GSED}). Following the procedure of avoiding ultraviolet divergences in Ref.~\cite{braaten1997}, the ground-state energy can be written as follows:
\begin{eqnarray}
	\frac{E_{\text{g}}}{V} & = & gn_{0}^{2}+g_{12}n_{0}^{2}+\frac{gn_{0}}{2V}\sum_{k\ne0}\left\{ f_{+}+f_{-}-\lim_{k\rightarrow\infty}\left(f_{-}+f_{+}\right)\right\} ,\label{OGS}
\end{eqnarray}
with
\begin{eqnarray}
\lim_{	k\rightarrow\infty}f_{+}& = & \lim_{	k\rightarrow\infty}\sqrt{\frac{{\rm Tr}\left(\left(\boldsymbol{\kappa}\boldsymbol{M}\right)^{2}\right)}{4\left(gn_{0}\right)^{2}}+\sqrt{\frac{{\rm Tr}\left(\left(\boldsymbol{\kappa}\boldsymbol{M}\right)^{2}\right)^{2}}{16\left(gn_{0}\right)^{4}}-\frac{\det\left(\boldsymbol{\kappa}\boldsymbol{M}\right)}{\left(gn_{0}\right)^{4}}}}\nonumber\\
&=&-\frac{\left(y-1\right)^{2}}{2\frac{\hbar k^{2}}{2mgn_{0}}\left(2z+1\right)}-\left(y-1\right)+\frac{\hbar k^{2}}{2mgn_{0}}\left(2z+1\right) ,\label{UDP}
\end{eqnarray}
and
\begin{eqnarray}
	\lim_{	k\rightarrow\infty}f_{-}& = & \lim_{	k\rightarrow\infty}\sqrt{\frac{{\rm Tr}\left(\left(\boldsymbol{\kappa}\boldsymbol{M}\right)^{2}\right)}{4\left(gn_{0}\right)^{2}}-\sqrt{\frac{{\rm Tr}\left(\left(\boldsymbol{\kappa}\boldsymbol{M}\right)^{2}\right)^{2}}{16\left(gn_{0}\right)^{4}}-\frac{\det\left(\boldsymbol{\kappa}\boldsymbol{M}\right)}{\left(gn_{0}\right)^{4}}}}\nonumber\\
	&=&\frac{\hbar k^{2}}{2mgn_{0}}+\left(1+y\right)-\frac{\left(y+1\right)^{2}}{2\frac{\hbar k^{2}}{2mgn_{0}}},\label{UDN}
\end{eqnarray}
then  Eq.~(\ref{OGS}) becomes
\begin{eqnarray}
	\frac{E_{\text{g}}}{V} & = & gn_{0}^{2}+g_{12}n_{0}^{2}+\frac{gn_{0}}{2V}\sum_{k\ne0}\left\{ f_{+}+f_{-}-\frac{\hbar^{2}k^{2}}{gn_{0}m}\left(1+z\right)-2+\frac{y^{2}+\left(y+1\right)^{2}z+1}{\left(2z+1\right)\frac{\hbar^{2}k^{2}}{2gn_{0}m}}\right\},\label{GS1}
\end{eqnarray}
which corresponds to Eq.~(\ref{GSED}) in main text.
\twocolumngrid
\bibliography{xyref}
\end{document}